%% file: RR-6621.tex
\documentclass[a4paper]{article}
\usepackage{RR}
\usepackage{hyperref}
\RRdate{Septembre 2008}

\RRauthor{
Bruno Guillaume\thanks{LORIA, INRIA Nancy Grand-Est (\url{Bruno.Guillaume@loria.fr})}%
  \and
Guy Perrier\thanks{LORIA, Universit\'e Nancy 2 (\url{Guy.Perrier@loria.fr})}%
}
\authorhead{B. Guillaume \& G. Perrier}
\RRtitle{Les Grammaires d'Interaction}
\RRetitle{Interaction Grammars}
\titlehead{Interaction Grammars}
\RRresume{Les grammaires d'interaction sont un formalisme grammatical
  bas\'e sur la notion de polarit\'e. Les polarit\'es expriment la
  sensibilit\'e aux ressources de la langue naturelle en distinguant les
  structures syntaxiques satur\'ees et insatur\'ees. La composition
  syntaxique peut \^etre vue comme une r\'eaction chimique control\'ee par
  la saturation des polarit\'es. Les grammaires sont exprim\'ees par un
  syst\`eme de contraintes utilisant la notion de description
  d'arbre. L'analyse syntaxique appara\^it alors comme un processus de
  construction de mod\`eles satisfaisant des crit\`eres de neutralit\'e et
  de minimalit\'e.}

\RRabstract{Interaction Grammar
  (IG) is a grammatical formalism based on the notion of
  polarity. Polarities express the resource sensitivity of natural
  languages by modelling the distinction between saturated and
  unsaturated syntactic structures. Syntactic composition is
  represented as a chemical reaction guided by the saturation of
  polarities. It is expressed in a model-theoretic framework where
  grammars are constraint systems using the notion of tree description
  and parsing appears as a process of building tree description models
  satisfying criteria of saturation and minimality.}
\RRmotcle{Formalisme grammatical, Grammaires cat\'egorielles, Polarit\'e, Description d'arbre}
\RRkeyword{Grammatical formalism, Categorial Grammar, Unification, Polarity, Tree description}
 \RRprojet{Calligramme}  
\RRtheme{\THSym} 
 \RCNancy 

\usepackage{graphicx,amssymb}
\usepackage[latin1]{inputenc}
\usepackage{covington}

\newcommand{\positive}{{\tt{->}}}
\newcommand{\negative}{{\tt{<-}}}
\newcommand{\neutral}{{\tt{=}}}
\newcommand{\virtual}{{$\sim$}}

\newcommand{\fneutr}[2]{{\tt{#1~=~#2}}}
\newcommand{\fpos}[2]{{\tt{#1~->~#2}}}
\newcommand{\fneg}[2]{{\tt{#1~<-~#2}}}
\newcommand{\feat}[1]{{\tt{#1}}}

\newcommand{\french}[1]{{\emph{``#1''}}}

\newcommand{\G}{{\cal G}}
\newcommand{\F}{{\cal F}}
\newcommand{\D}{{\cal D}}
\newcommand{\I}{{\cal I}}
\newcommand{\rI}{{\cal I}^{-1}}
\newcommand{\T}{{\cal T}}
\renewcommand{\P}{{\cal P}}
\renewcommand{\S}{{\cal S}}
\newcommand{\NT}{{\cal NT}}
\newcommand{\NP}{{\cal NP}}
\newcommand{\nl}{{\hfil\\}}
\renewcommand{\ss}{{\prec\!\!\prec}}

\newcommand{\agram}{{${}^\ast$}}

\newcommand{\leopar}{{\sc Leopar}}
\newcommand{\lefff}{Le{\em fff}}
\begin{document}
\RRNo{6621}
\makeRR

\input{main_rr}

\end{document}

%% file: main_rr.tex
\section*{Introduction}

Interaction Grammar (IG) is a grammatical formalism based on an old idea
of O.~Jespersen~\cite{Jes37}, L.~Tesni\`ere~\cite{Tes34} and
K.~Adjukiewicz~\cite{Adj35}: a sentence is viewed as a molecule with its words
as the atoms; every word is equipped with a valence which expresses its capacity
of interaction with other words, so that syntactic composition appears as a
chemical reaction.

The first grammatical formalism that exploited this idea was Categorial Grammar
(CG)~\cite{Ret00}. In CG, constituents are equipped with types, which express
their interaction ability in terms of syntactic categories.
A way of highlighting this originality is to use polarities: syntactic types can
be represented by partially specified syntactic trees, which are decorated with
polarities that express a property of non saturation; a positive node represents
an available grammatical constituent whereas a negative node represents an
expected grammatical constituent; negative nodes tend to merge with positive
nodes of the same type and this mechanism of neutralization between opposite
polarities drives the composition of syntactic trees to produce saturated trees
in which all polarities have been neutralized.

The notion of polarity in this sense was not used explicitly in computational
linguistics until recently. To our knowledge, A.~Nasr was the first to propose a
formalism using polarized structures~\cite{Nas95}. Then, nearly at the same
time, R.~Muskens~\cite{Mus98}, D.~Duchier and S.~Thater~\cite{Duc99}, and
G.~Perrier~\cite{Per00} proposed grammatical formalisms using polarities. The
latter was a first version of IG, presented in the framework of linear
logic. This version, which covers only the syntax of natural languages, was
extended to the semantics of natural languages~\cite{Per05}.
Then, S.~Kahane showed that all well known formalisms (CFG, TAG, HPSG, LFG) can
be viewed as polarized formalisms~\cite{Kah06}.
Unlike the previous approaches, polarities are used in a non monotonous way in 
Minimalist Grammar (MG). E.~Stabler~\cite{Sta97} proposes a formalization of MG
which highlights this. Polarities are associated with syntactic features to
control movement inside syntactic structures: strong features are used to drive
the movement of phonetic forms (overt movement) and weak features are used to
drive the movement of logical forms (covert movement).

With IG, we highlighted the fundamental mechanism of saturation between
polarities underlying CG in a more refined way, because polarities are attached
to the features used to describe constituents and not to the constituents
themselves --- but the essential difference lies in the change of framework: CG
are usually formalized in a generative deductive framework, the heart of which
is the Lambek Calculus~\cite{Lam58}, whereas IG is formalized in a
model-theoretic framework. A particular interaction grammar appears as a set of
constraints, and parsing a sentence with such a grammar reduces to solving a
constraint satisfaction problem. G.~K.~Pullum and B.~C.~Scholz highlighted the
advantages of this change of framework~\cite{Pul01}. Here, we are especially
interested in some of these advantages:
\begin{itemize}
\item syntactic objects are tree descriptions which combine independent
elementary properties in a very flexible way to represent families of syntactic
trees;
\item underspecification can be represented in a natural way by tree
descriptions;
\item partially well-formed sentences have a syntactic representation in the
sense that, even if they have no complete parse trees, they can be characterized
by tree descriptions.
\end{itemize}

The notion of tree description, which is central in this approach, was
introduced by M.~Marcus, D.~Hindle and M.~Fleck to reduce
non-determinism in the parsing of natural languages~\cite{Marc83}. It
was used again by K.~Vijay-Shanker to represent the adjoining
operation of TAG in a monotonous way~\cite{Vij92}. Then, it was
studied systematically from a mathematical point of view~\cite{Rog92}
and it gave rise to new grammatical formalisms~\cite{Kal99,Ram01}.

If model theory provides a declarative framework for IG, polarities
provide a step by step operational method to build models of tree
descriptions: partially specified trees are superposed\footnote{As no
  standard term exists, we use the term ``superposition'' to name the
  operation where two trees are combined by merging some nodes of the
  first one with nodes of the second one.}  under the control of
polarities; some nodes are merged in order to saturate their
polarities and the process ends when all polarities are saturated. At
that time, the resulting description represents a completely specified
syntactic tree. The ability of the formalism to superpose trees is
very important for its expressiveness. Moreover, the control of
superposition by polarities is interesting for computational
efficiency.

In natural languages, syntax is a way to access semantics and a linguistic
formalism worthy of the name must take this idea into account. If the goal of
the article is to give a formal presentation of IG which focuses on the
syntactic level of natural languages, the formalism is designed in such a way
that various formalizations of semantics can be plugged into IG. The reader can
find a first proposal in~\cite{Per05}.

An important concern with IG is to provide a realistic formalism,
which can be experimented parsing actual corpora. In order to combine
the theoretical development of the formalism with experimentation, we
have designed a parser, \leopar, based on IG~\cite{Bon03}.  If a
relatively efficient parser is a first condition to get a realistic
formalism, a second condition is to be able to build large coverage
grammars and lexicons. With an appropriate tool, XMG~\cite{Duc04}, we
have built a French interaction grammar with a relatively large
coverage~\cite{Per2007}. This grammar is designed in such a way that
it can be linked with a lexicon independent of any formalism.  Since
our purpose in this article is to present the formal aspects of IG, we
will not dwell on the experimental side.

The layout of the paper is as follows:
\begin{itemize}
\item Section~\ref{sec:global} gives an intuitive view of the main IG features
(polarities, superposition and underspecification) through significant examples.
\item Section~\ref{sec:tree_desc} presents the syntax of the language used to
represent polarized tree descriptions, the basic  objects of the formalism.
\item Section~\ref{sec:models} explains how syntactic parse trees are
  related to polarized tree descriptions with the notion of minimal
  and saturated model.
\item In section~\ref{sec:expressivity}, we illustrate the expressivity of IG
with various linguistic phenomena.
\item In section~\ref{sec:comparison}, we compare IG with the most closely
related formalisms.
\item Section~\ref{sec:leopar} briefly presents the computational aspects of IG
through their implementation in the \leopar\ parser, which works with a
relatively large French interaction grammar.
\end{itemize}

\section{The main features of Interaction Grammars}
\label{sec:global}
The aim of this section is to give informally, through examples, an
overview of the key features of IG.

\subsection{A basic example}

\subsubsection{Syntactic tree}
\begin{figure}[!h] 
    \includegraphics[width=6cm]{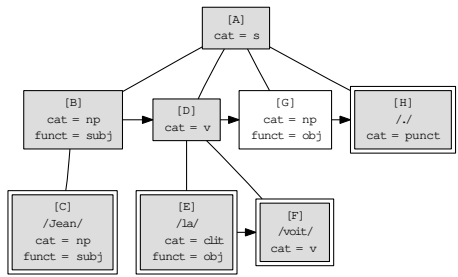} 
    \caption{Syntactic tree for the sentence \french{Jean la voit.}}
    \label{model} 
\end{figure}

In IG, the parsing output of a sentence is an ordered tree where nodes
represent syntactic constituents described by feature structures. An
example of syntactic tree for sentence (\ref{ex:jean_la_voit}) is
shown in Figure~\ref{model}\footnote{To increase readability, only a
  part of the feature structures is shown in the figures; many other
  features (gender, number, mood, \ldots) are used in practice. In the
  following, we only show relevant features in figures.}.

\begin{examples}
\item \label{ex:jean_la_voit}
  \gll Jean la voit.
  John it sees.
  \glt `John sees it.'
  \glend
\end{examples}

Each leaf of the tree carries a phonological
form which is a string that can be empty (written~$\epsilon$): in our
example, \french{Jean} in node [C], \french{la} in [E], \french{voit}
in [F], $\epsilon$ in [G] and \french{.} in [H]). The {\em
  phonological projection} of a tree is the left to right reading of
the phonological forms of its leaves
(\french{Jean}$\cdot$\french{la}$\cdot$\french{voit}$\cdot\epsilon\cdot$\french{.}
= \french{Jean la voit.} in the example).

\subsubsection{Initial tree descriptions}
\begin{figure}[!h] 
    \includegraphics[width=12cm]{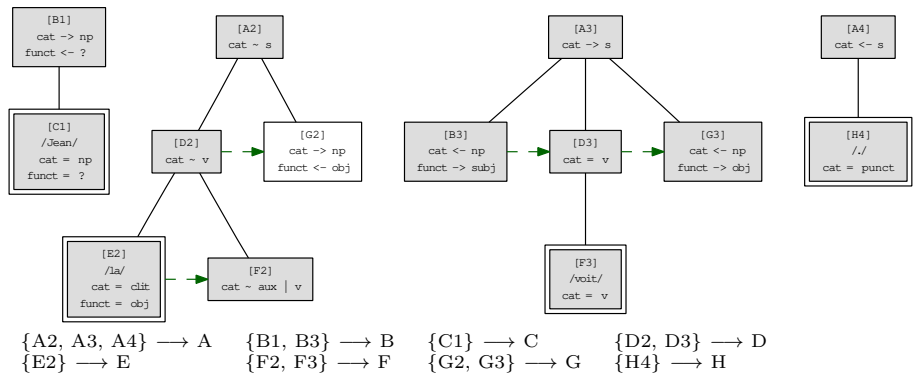} 
{\scriptsize
\begin{tabular}{llll}
\{A2, A3, A4\} $\longrightarrow$ A & \{B1, B3\} $\longrightarrow$ B & \{C1\} $\longrightarrow$ C & \{D2, D3\} $\longrightarrow$ D \\ 
\{E2\} $\longrightarrow$ E & \{F2, F3\} $\longrightarrow$ F & \{G2, G3\} $\longrightarrow$ G & \{H4\} $\longrightarrow$ H \\
\end{tabular}
}

    \caption{IPTDs and interpretation function for the sentence \french{Jean la voit.}}
    \label{iptds} 
\end{figure}

The elementary syntactic structures are {\em initial polarized tree
  descriptions} (written IPTDs in the following). Figure~\ref{iptds}
shows the four IPTDs used to build the syntactic tree in
Figure~\ref{model}. A syntactic tree is said to be a model of a set of
IPTDs if each node of the syntactic tree interprets some nodes of the
IPTDs and this tree satisfies saturation and minimality
constraints. For our example, the interpretation function is also given
in Figure~\ref{iptds}.

IPTDs are underspecified trees: for instance, in Figure~\ref{iptds},
the precedence relation between nodes [D2] and [G2] is large: [D2]
must be to the left of [G2] but any number of intermediate nodes
between [D2] and [G2] are allowed in the final tree model.

Moreover, IPTDs contain features with polarities acting as
constraints. A positive (written~\positive) polarity must be
associated with a compatible negative (written~\negative) one: in the
example, when building the model, the positive feature \fpos{cat}{s}
of node [A3] is associated with the negative feature \fneg{cat}{s} of
node [A4].

\subsubsection{Tree descriptions}
A more general notion of tree description is not strictly needed in the
formalism definition, however this notion is useful to represent
partial parses of sentence and to consider atomic steps in parsing
process. These polarized tree descriptions (PTDs) are formally described in
the next section.

\subsection{Polarized features to control syntactic composition}
The notion of polarity represents the core of the IG formalism. 

\subsubsection{Positive and negative polarities}
Like in categorial grammars, resources can be identified as available
(positive polarity) or needed (negative polarity). Each positive or
negative feature must be neutralized by a dual polarity when the model
is built. A polarity which is either positive or negative is said to
be {\em active}.

This mechanism is intensively used. It is used similarly as in CG, for
instance, to control the interactions of:
\begin{itemize}
\item a determiner with a noun;
\item a preposition with a noun phrase;
\item a verb, a predicate noun or adjective with its arguments defined in the
subcategorization frame.
\end{itemize}

But polarities are also used in a more specific manner in IG to deal
with other kinds of interactions. For instance:
\begin{itemize}
\item to handle pairs of grammatical words like {\em ne/pas}, \ldots (see below
  subsection~\ref{pairing});
\item to manage interaction of punctuation with other constructions in the
sentence;
\item to link a reflexive pronoun {\em se} with the reflexive construction of 
verbs;
\item to manage interaction between auxiliaries and past participles.
\end{itemize}

\subsubsection{Virtual polarities}
Recently, a third kind of polarity was added which is called {\em
  virtual} (written~\virtual). A feature with a virtual polarity must
be combined with some other compatible feature which has a polarity
different from~\virtual. It gives more flexibility to express
constraints on the context in which a node can appear. Virtual
polarities are used, for instance:
\begin{itemize}
\item to describe interaction between a modifier and the modified
  constituent (adverb, adjective, \ldots), see
  subsection~\ref{modifier} for an example;
\item to express context constraints on nodes around the active part
  of a description; it allows for a control on the superposition
  mechanism: in Figure~\ref{iptds}, the three nodes [A2], [D2] and
  [F2] with virtual {\tt cat} polarities describe the context in which
  the clitic \french{la} must be used; this IPTD requires that three
  other non-virtual nodes compatible with [A2], [D2] and [F2] exist in
  some other IPTDs; in our example, non-virtual nodes [A3], [D3] and
  [F3] are given by the verb. This mechanism handles the constraint on
  the French clitic \french{la}. It comes before the verb (node [E2]
  before node [F2]) but contributes with an object function (node [G2]
  after node [F2] because the canonical position of French direct
  object in on the right of the verb).
\end{itemize}

\subsubsection{Polarities at the feature level}
A difference with respect to other formalisms using polarities is
that, in IG, polarities are attached to features rather than to
nodes. It is then possible to use polarities for several different
features to control different types of positive/negative pairing (for
instance in our grammar, the feature {\tt mood} is polarized in the
auxiliaries/past participles interaction; the feature {\tt neg} is
polarized in the interaction of the two pieces of negation).

Hence with polarities at the feature level, the same syntactic
constituent can interact more than once with its environment through
several feature neutralizations.

One of the typical usage of such interactions, that implies more than
two nodes is subject inversion. In French, in some specific cases the
subject can be put after the verb (sentences (\ref{inv_subj_rel_1}),
(\ref{inv_subj_rel_2}) and (\ref{inv_subj_rel_3})). However,
uncontrolled subject inversion would lead to over-generation. A
solution is to use two different interactions: between the subject and
the verb on one hand; and on the other hand between the subject and
some other word which is specific to the construction where the
subject can be postponed.

\begin{examples}
\item \label{inv_subj_rel_1}
  \gll Jean qu'aime Marie vient.
  John that loves Mary comes.
  \glt `John that Mary loves comes.'
  \glend
\item \label{inv_subj_rel_2} 
  \gll Aujourd'hui commence le printemps.
  Today begins the spring.
  \glt `Today begins the spring.'
  \glend
\item \label{inv_subj_rel_3}
  \gll Que {} mange Jean~?
  What does eat John?
  \glt `What does John eat?'
  \glend
\end{examples}

\begin{figure}[!h]
  \begin{center}
    \includegraphics[width=13cm]{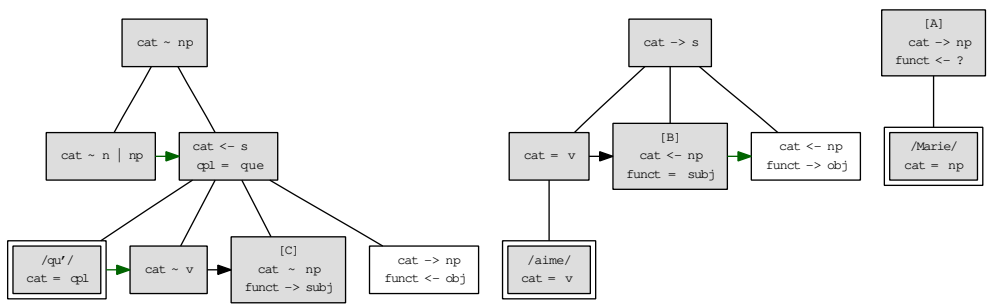} 

    \caption{IPTDs for the sequence of words \french{qu'aime
        Marie} before superposition}
    \label{qu_aime_marie}
  \end{center} 
\end{figure}

\begin{figure}[!h]
  \begin{center}
    \includegraphics[width=8cm]{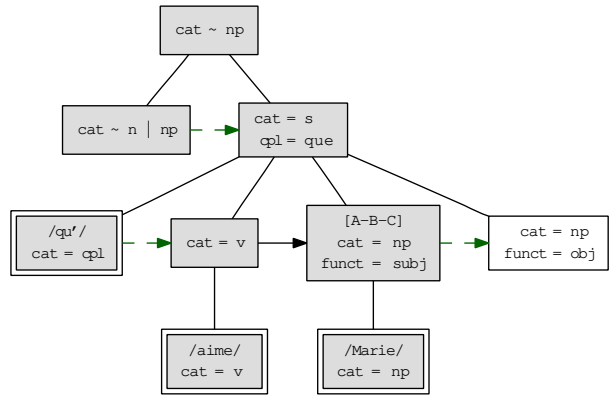} 

    \caption{PTD for the sequence of words \french{qu'aime
        Marie} after superposition}
    \label{qu_aime_marie_final}
  \end{center} 
\end{figure}

In the sentence~(\ref{inv_subj_rel_1}), the subject \french{Marie} of
the verb \french{aime} can be postponed because it is in a relative
clause introduced by the object relative pronoun \french{que}. Hence,
in the noun phrase \french{Jean qu'aime Marie} (see
figure~\ref{qu_aime_marie}), the proper noun \french{Marie} interacts
both with the verb \french{aime} (neutralization of the features
\fpos{cat}{np} in [A] and \fneg{cat}{np} in [B]) and with
the relative pronoun \french{qu' } (neutralization of the features
\fneg{funct}{?} in [A] and
\fpos{funct}{subj} in [C]). Figure~\ref{qu_aime_marie_final} gives the PTD
after superposition.

\subsection{Tree superposition as a flexible way of realizing syntactic
composition}
For the grammatical formalisms that are based on trees (the most
simple formalism of this type is Context Free Grammar), the mechanism
of syntactic composition often reduces to substitution: a leaf
$L$ of a first tree merges with the root $R$ of a second
tree. In this way, constraints on the composition of both trees are
localized at the nodes $R$ and $L$. They cannot say anything
about the environment of both nodes.

The TAG formalism offers a more sophisticated operation,
\emph{adjunction}, but this operation is also limited in expressing
constraints on syntactic composition: instead of merging two nodes, we
merge two pairs of nodes. A node $N$ splits into an up node $N_{up}$
and down node $N_{down}$, which respectively merge with the root $R$
and the foot $F$ of the auxiliary tree. Constraints on syntactic
composition is now localized on three nodes $N$, $R$ and $F$.

In IG, the syntactic composition is much more flexible: we can merge
any two nodes (in the same PTD or in two different ones). Then, the
propagation of the constraints related to each PTD entails a partial
superposition of the two tree structures around the two nodes. In this
way, we can express constraints on the environment of a node.

\begin{figure}[!h]
\begin{center}
\includegraphics[width=8cm]{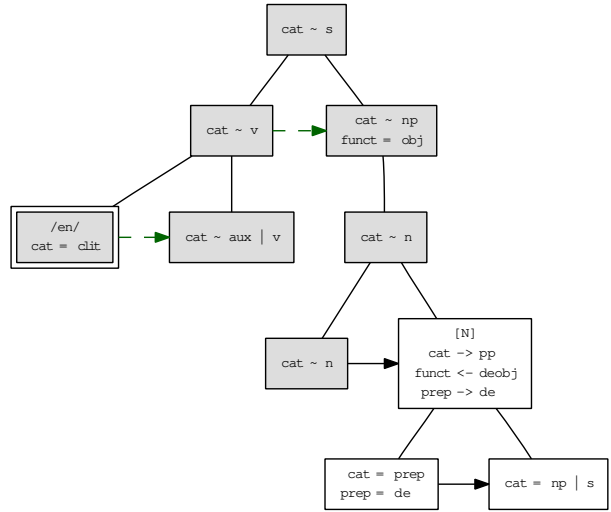}
\caption{IPTD representing the syntax of the clitic \french{en}} 
\label{en}
\end{center}
\end{figure}

\begin{figure}[!h]
\begin{center}
\includegraphics[width=10cm]{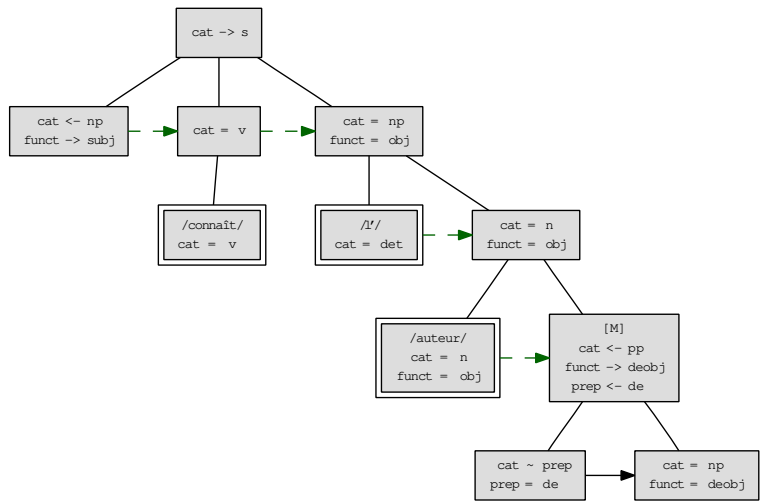}
\caption{PTD representing the syntax of the phrase \french{connaît l'auteur}} 
\label{connait_l_auteur}
\end{center}
\end{figure}

\begin{examples}
\item \label{ex:auteur}
  \gll Jean en connaît l'auteur.
  John {of it} knows the author.
  \glt `John knows the author of it.'
  \glend
\end{examples}

Let us consider the sentence (\ref{ex:auteur}). The clitic pronoun
\french{en} provides the object \french{auteur} of the verb
\french{connaît} with a noun complement. Our French lexicon gives the
IPTD of Figure~\ref{en} to represent the syntax of this usage of the
clitic pronoun \french{en}.  In this IPTD, the node [N] with feature
\fpos{prep}{de} represents the trace of the preposition phrase
represented by the clitic \french{en} as a sub-constituent of the
object of the verb.  Figure~\ref{connait_l_auteur} shows a PTD
resulting from the (partial) parsing of \french{connaît l'auteur}. In
this PTD, the node [M] with feature \fneg{prep}{de} represents the
noun complement that is expected by the noun \french{auteur}.

\begin{figure}[!h]
\begin{center}
\includegraphics[width=11cm]{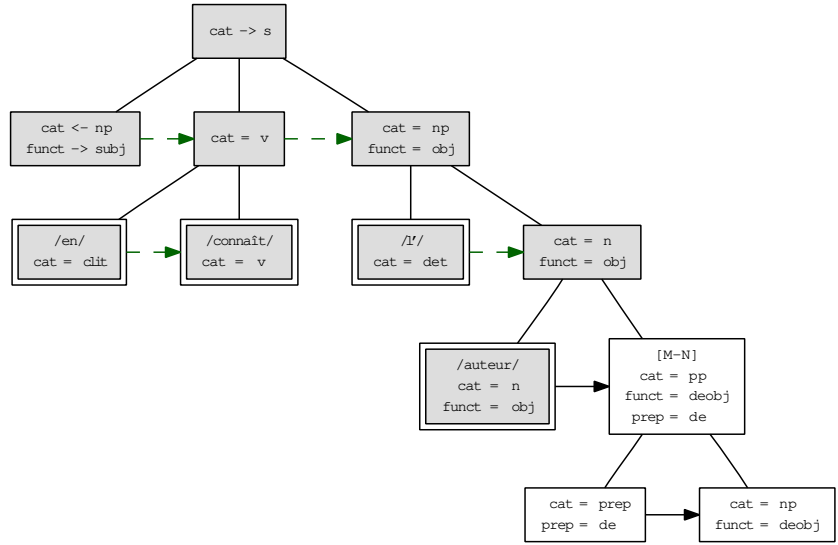}
\caption{PTD representing the syntax of the phrase \french{en connaît
l'auteur}} 
\label{en_connait_l_auteur}
\end{center}
\end{figure}

Now, when we compose \french{en} with \french{connaît l'auteur}
(i.e. tree descriptions of Figures~\ref{en}
and~\ref{connait_l_auteur}), nodes [N] and [M] have to be merged in
order to neutralize their features \feat{cat}, \feat{funct} and
\feat{prep}. By propagating tree well-formedness and polarity
constraints, the merging of [N] and [M] entails the partial
superposition (Figure~\ref{en_connait_l_auteur}) of the two PTDs. Note
that there are 9~atomic operations of node merging during this
composition.

\subsection{Underspecified structures}

With IG, both dominance and precedence relations can be underspecified:
an IPTD can constrain a relation between two nodes without
restricting the distance between the nodes in the
model. Underspecified relations, combined with tree superposition,
increase the flexibility of the formalism: it is possible to give more
general constraints on the context of a node.

Underspecification on dominance relation makes it possible to express
general properties on unbounded dependencies. For instance, the
relative pronoun \french{que} can introduce an unbounded dependency
between its antecedent and a verb which has this antecedent as object
of adjectival complement: sentences~(\ref{under_dom_1}) and
(\ref{under_dom_2})\footnote{The symbol $\square$ indicates the original place of the extracted argument.}.

\begin{examples}
\item \label{under_dom_1}
 \gll Jean {\bf que} Marie aime $\square$ dort.
  John that Mary loves {} sleeps
  \glt
  \glend
\item \label{under_dom_2}
  \gll Jean {\bf que} Pierre croit que Marie aime $\square$ dort.
  John that Peter thinks that Mary loves {} sleeps
  \glt 
  \glend
\end{examples}

\begin{figure}[!h]
\begin{center}
\includegraphics[width=8cm]{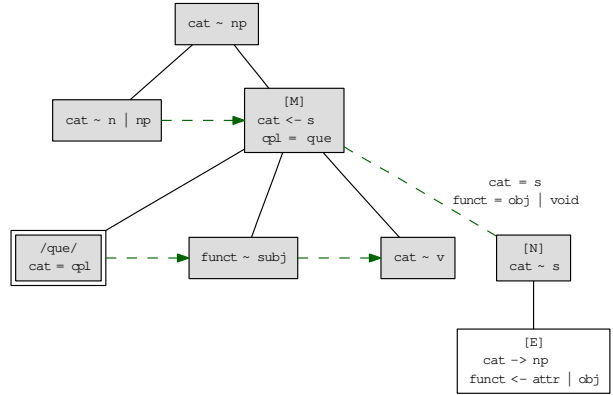}
\caption{IPTD for the relative pronoun {\em que}}
\label{que}
\end{center}
\end{figure}

\begin{figure}[!h]
\begin{center}
\includegraphics[width=9cm]{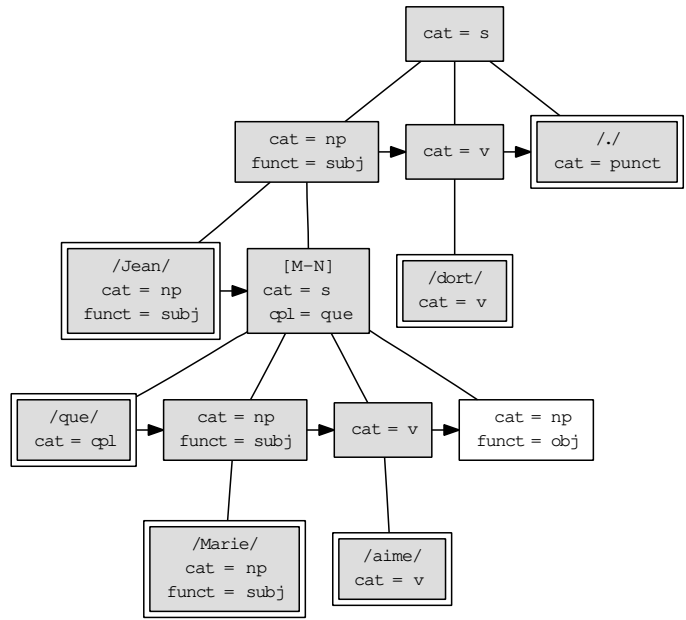}
\caption{Syntactic tree for the sentence (\ref{under_dom_1})} 
\label{fig_under_dom_small}
\end{center}
\end{figure}

\begin{figure}[!h]
\begin{center}
\includegraphics[width=12cm]{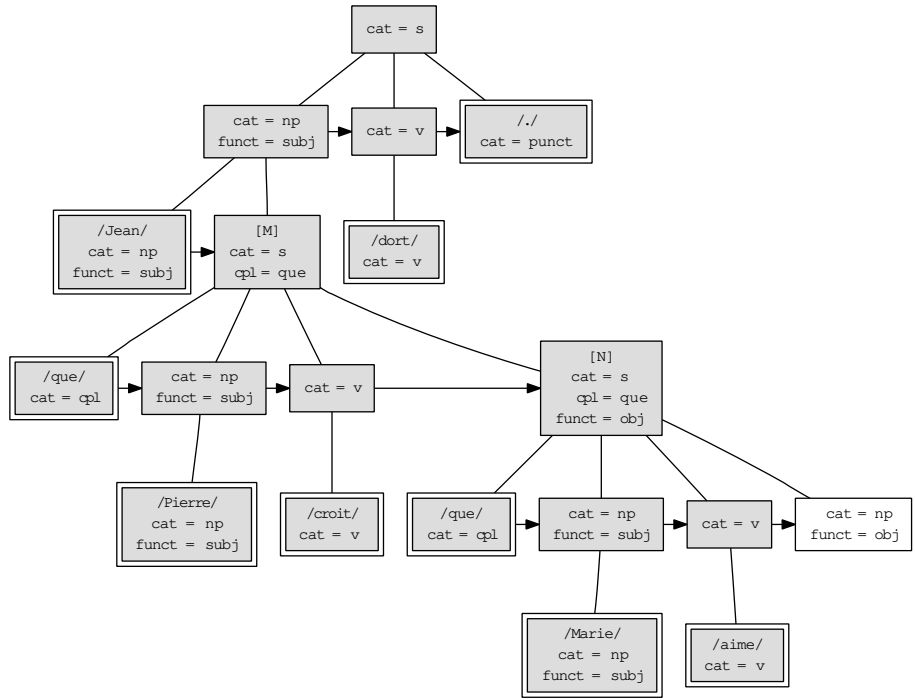}
\caption{Syntactic tree for the sentence (\ref{under_dom_2})} 
\label{fig_under_dom_large}
\end{center}
\end{figure}

Figure~\ref{que} provides an IPTD to model this use of
\french{que}. An empty node~[E] represents the trace of an object or
an adjectival phrase; [N] represents the clause in which the trace is
a direct constituent and [M] represents the relative clause introduced
by the relative pronoun \french{que}. [N] can be embedded at any depth
in [M], which is expressed by an underspecified dominance
relation. Figure~\ref{fig_under_dom_small} shows a model for the
sentence (\ref{under_dom_1}) in which the relation is realized by
merging [M] and [N], whereas Figure~\ref{fig_under_dom_large} shows a
model for the sentence (\ref{under_dom_2}) in which the relation is
realized by an immediate dominance relation.

In order to deal with island constraints, large dominances need to be
controlled. In IG, this is possible with the notion of filtering feature
structures. A filtering feature structure is a polarized feature
structure where all polarities are neutral. A large dominance $M >^*
N$ labelled with a filtering feature structure $\psi$ means that node $M$
must dominate $N$ in the model and that each node along the path from
$M$ to $N$ in this model must be compatible with $\psi$. For instance,
in Figure~\ref{que}, such a filter is used to avoid extraction through
nodes that are not of category {\tt s}.

\begin{figure}[!h]
\begin{center}
\includegraphics[width=9cm]{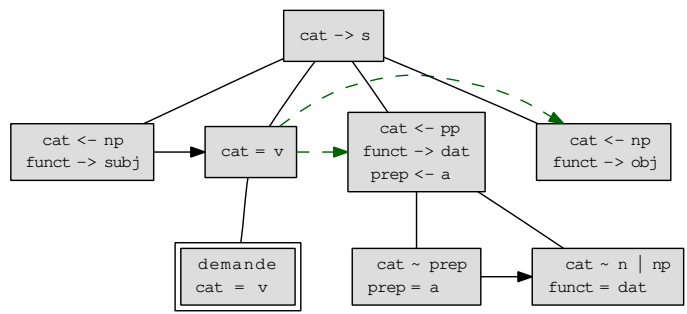}
\caption{IPTD for the verb \french{demande}} 
\label{demande}
\end{center}
\end{figure}

With underspecification on precedence relation, it is possible to
describe a free ordering of some arguments. For instance, both
sentences (\ref{under_prec_1}) and (\ref{under_prec_2}) can be parsed using the
same IPTD (Figure~\ref{demande}) for the word \french{demande}.
	
\begin{examples}
\label{under_dom}
\item \label{under_prec_1}
 \gll Jean demande une invitation à Marie.
  John asks an invitation to Mary.
  \glt `John asks an invitation to Mary.'
  \glend
\item \label{under_prec_2}
  \gll Jean demande à Marie une invitation.
  John asks {} Mary an invitation
  \glt `John asks Mary an invitation.'
  \glend
\end{examples}

\section{Formal definitions}
\label{sec:tree_desc}

This section is dedicated to formal definitions of IG. We
define in turn:
\begin{itemize}
\item {\em syntactic trees}: the final syntactic structures in the
  parsing process;
\item {\em initial polarized tree descriptions} (IPTDs): the initial syntactic
  structures that are associated to words at the beginning of the
  parsing process; PTDs are also defined as a generalization of IPTDs;
\item the notion of {\em model} which links IPTDs and syntactic
  trees.
\end{itemize}

\subsection{Syntactic trees}

\subsubsection{Features}
Features are built relatively to a feature signature. A {\em feature
  signature} is defined by:
\begin{itemize}
\item a finite set $\F$ of constants called {\em feature names};
\item for each feature name in $\F$ a finite set $\D_f$ of constants
  called {\em atomic values}.
\end{itemize}

A {\em feature} is a couple $(f,v)$ where $f \in \F$ and $v \in \D_f$
and a {\em feature structure} is a set of features with different
feature names.

\subsubsection{Syntactic trees}
A {\em syntactic tree} is a totally ordered tree where:
\begin{itemize}
\item each node carries a feature structure,
\item each leaf carries a string (which can be the empty string
  written $\epsilon$) called {\em phonological form}.
\end{itemize}

In syntactic trees, parenthood relation is written $M \gg N$ (this
means that $M$ is the mother node of $N$), immediate precedence
between sisters is written $M \ss N$ (this means that $M$ and $N$ have
the same mother and that $M$ is just before $N$ in the sisters
ordering)\footnote{We use double symbols to avoid confusion with
  relations that are defined later for IPTDs.}. We also use the
notation $M \gg [N_1,\ldots,N_k]$ when the set of daughters of $M$ is
the ordered list $[N_1,\ldots,N_k]$.

Let $\gg^*$ denote the reflexive and transitive closure of $\gg$.
If $M \gg^* M'$ then we call $path(M,M')$ the list of nodes from $M$ to
$M'$:

$$path(M,M') = \{N_i\}_{1 \leq i \leq n} {\rm \ such\ that\ } 
\left\lbrace
  \begin{array}{l}
    N_1 = M \\
    N_i \gg N_{i+1} \rm{\ for\ } {1 \leq i < n} \rm \\
    N_n = M
  \end{array}
\right.
$$

We define the {\em phonological projection} $PP(M)$ of a node $M$ to
be the list of non-empty strings built with the left to right reading of the
phonological forms in the subtree rooted by $M$:

\begin{itemize}
\item if $M \gg []$ (i.e. $M$ is a leaf) and the phonological form of $M$ is
$\epsilon$ then $PP(M) = []$,
\item if $M \gg []$ and the phonological form of $M$ is the non-empty string
$phon$ then $PP(M) = [phon]$,
\item if $M \gg [N_1,\ldots,N_k]$ then $PP(M) = PP(N_1) \circ \ldots \circ
PP(N_k)$ (where $\circ$ is the concatenation of lists).
\end{itemize}

The phonological projection of a syntactic tree is the phonological
projection of its root.

We conclude here with a remark. The fact that syntactic trees are
completely ordered trees can sometimes produce unwanted effects. For
instance, when a node has several empty daughters, it may be not
relevant to consider the relative order of these nodes. In
sentences~(\ref{under_prec_1}) and (\ref{under_prec_2}), the verb
\french{demander} with a direct object and a dative does not impose
any order between arguments. When the two arguments are realized as
clitics in sentence (\ref{clitic}), the relative order of clitics is
fixed but there are two models with different ordering on empty nodes
corresponding to the two arguments.

\begin{examples}
\item \label{clitic} 
  \gll Jean la lui demande.
  John it {to her} asks.
  \glt 'John asks it to her.'
 \glend
\end{examples}

In order to avoid this problem, it is possible to define an
equivalence relation that identifies the two models of the
sentence (\ref{clitic}). We will not detail this relation in
this article.

\subsection{Polarized tree descriptions}
\subsubsection{Polarities}
Polarities are heavily used in IG to take into account the resource
sensitivity of natural languages. Furthermore, the parsing process
strongly relies on these polarities.

The current IG formalism uses four polarities:

\begin{itemize}
\item {\em positive} (written \positive): a feature with a positive
  polarity describes an available resource;
\item {\em negative} (written \negative): a feature with a negative
  polarity describes a needed resource;
\item {\em virtual} (written \virtual): a feature with a virtual
  polarity is waiting for unification with another non-virtual one;
  virtual polarities are used for expressing constraints on the
  context in which an IPTD can be inserted;
\item {\em neutral} (written \neutral): a feature with a neutral
  polarity is not concerned by the resource management: it acts like a
  filter in case of unification; but unification is not required.
\end{itemize}

A multiset of polarities is said to be {\em globally saturated}:
\begin{itemize}
\item if it contains exactly one positive and one negative polarity;
\item or if it contains no positive, no negative and a least one neutral
polarity.
\end{itemize}

\subsubsection{Polarized features}
Whereas features in final syntactic trees are defined by a couple
name value, in the tree description a polarity is attached to
each feature and the feature values can be underspecified (with a
disjunction of atomic values).

Hence, {\em polarized features} are now defined by triples of:
\begin{itemize} 
\item a feature name $f$ taken from $\F$,
\item a polarity,
\item a feature value which is a disjunction of atomic values taken
  from $\D_f$; a feature value is written as a list of atomic
  values separated by the pipe symbol {\tt |}; the question mark symbol
  {\tt ?} denotes the disjunction of all values in $\D_f$.
\end{itemize}

A polarized feature is written as the concatenation of these three
components (for instance \fpos{cat}{np|pp}, \fneg{funct}{?} are
polarized features).

It is also possible to give additional constraints on feature values
with co-references. A co-reference is noted with {\tt <}$i${\tt >}; for
instance \fneutr{mood}{<2>~ind|subj} is a co-referenced feature.

\subsubsection{Polarized feature structures}
A {\em polarized feature structure} is a set of polarized features
with different feature names.

\subsubsection{Filtering feature structures}
{\em Filtering feature structures} are used to represent constraints on
underspecified dominances. A {\em filtering feature structure} is a polarized
feature structure where all polarities are neutral.

The constraints on underspecified dominances are stated in terms of
{\em compatibility}. A feature structure $\varphi$ is said to be {\em
  compatible} with a filtering feature structure $\Psi$ (notation
$\varphi \triangleleft \Psi$) if, for each feature name $f$ defined in
both structures, the atomic value associated with $f$ in $\varphi$ is
included in the disjunction associated with $f$ in $\Psi$.

\subsubsection{Polarized nodes}
A {\em polarized node} is described by:
\begin{itemize}
\item a polarized feature structure;
\item a node type.
\end{itemize}

Node types express constraints on the phonological projection of nodes in the
model. Each node has one of these four types:

\begin{itemize}
\item {\bf anchor} with an associated phonological form (a non-empty
  string): the image of an anchor must be a leaf of the tree model
  (anchors are drawn with a double border in figures);
\item {\bf full}: a full node must have an image with a non-empty
  phonological projection;
\item {\bf empty}: an empty node must have an image with an empty
  phonological projection (empty nodes are drawn with white background
  in figures);
\item {\bf default}: a default node has no constraint on its phonological
projection.
\end{itemize}





\subsubsection{Polarized tree descriptions}
We consider four types of relation between nodes in our tree
descriptions: 

\begin{description}
\item [dominance]\nl The relation $M > N$ constrains the image of $M$
  to be the mother of the image of $N$. In such a relation it can also
  be imposed that $N$ is the leftmost (resp. rightmost) daughter of
  $M$: we write $M > \bullet N$ (resp. $M > N\bullet$). Finally, an
  arity constraint can be expressed on the set of daughters of a node:
  $M > \{N_1, \ldots, N_k\}$ imposes that the image of $M$ in the
  model has exactly $k$ daughters that are images of the $N_i$ (this
  arity constraint does not impose any order on the $k$ daughters of
  the node $M$).
\item [large dominance]\nl $M >^* N$ constrains the image of $N$ to be
  in the subtree rooted at the image of $M$\footnote{Note that the
    symbol $>^*$ is another relation which is not defined as the
    reflexive and transitive closure of the relation $>$. The same
    remark applies to relations $\prec^+$ and $\prec$ defined below.}.
  A large dominance can also carry an additional constraint on the nodes
  that are on the path from $M$ to $N$ in the model: $M >^*_{\Psi} N$
  (where $\Psi$ is a filtering feature structure) constrains that the
  image of $N$ is in the subtree rooted at the image of $M$ and
  that each node along the path between the two images carries a
  feature structure which is compatible with $\Psi$.
\item [precedence]\nl $M \prec N$ constrains the images of the two
  nodes to be daughters of the same node in the model and the image of
  $M$ to be the immediate left sister of the image of $N$;
\item [large precedence]\nl $M \prec^+ N$ constrains the images of the two
  nodes to be daughters of the same node in the model and the image of $M$
  to precede the image of $N$ in the ordered tree; this precedence is
  strict, hence the two images have to be different.
\end{description}

A {\em polarized tree description} (PTD) is defined by:

\begin{itemize}
\item a set of polarized nodes;
\item a set of relations on these nodes which verifies the condition:
  if $N_1 \prec N_2$ or $N_1 \prec^+ N_2$ then there is a node $M$
  such that $M > N_1$ and $M > N_2$.
  
  Note that this condition imposes that $N_1$ and $N_2$ have the same
  mother in the IPTD and not only in the model.
\end{itemize}

\subsubsection{Initial polarized tree descriptions}
IPTDs are elementary structures that are linked with words in the
grammar; an IPTD is a PTD which verifies the additional constraint:
the relation $> \cup >^*$ defines a tree structure on the nodes, this
implies connexity and the fact that except the root node, all other
nodes $N$ have exactly either one mother node $M$ ($M > N$) or one
ancestor node $M$ ($M >^* N$ or $M >^*_{\Psi} N$).

\section{Syntactic trees as models of IPTDs}
\label{sec:models}

The aim of this section is to describe precisely the link between
IPTDs and syntactic trees.

\subsection{Syntactic trees as models of set of IPTDs}


Let $\G$ be an interaction grammar. A syntactic tree $\T$ is a model
of a multiset of IPTDs $\P = \{\P_i\}_{1 \leq i \leq k}$ if there is
{\em an interpretation function} $\I$ from the nodes $\NP$ of the multiset
$\P$ to nodes $\NT$ of the syntactic tree $\T$ such that:

\begin{description}
\item [Dominance adequacy]\nl\vspace{-1.6em}
  \begin{itemize}
    \item if $M, N \in \NP$ and $M > N$ then $\I(M) \gg \I(N)$.
  \end{itemize}
\item [Large dominance adequacy]\nl\vspace{-1.6em}
  \begin{itemize}
  \item if $M, N \in \NP$ and $M >^* N$ then $\I(M) \gg^* \I(N)$.
  \item if $M, N \in \NP$ and $M >^*_{\Psi} N$ then $\I(M) \gg^*
    \I(N)$ and for each node $P$ in $path (\I(M),\I(N))$, $\varphi(P)
    \triangleleft \Psi$.
  \end{itemize}
\item [Precedence adequacy]\nl\vspace{-1.6em} 
  \begin{itemize} 
  \item if $M, N \in \NP$ and $M \prec N$ then $\I(M) \ss \I(N)$.
  \end{itemize}
\item [Large precedence adequacy]\nl\vspace{-1.6em} 
  \begin{itemize} 
  \item if $M, N \in \NP$ and $M \prec^+ N$ then $\I(M) \ss^+ \I(N)$.
  \end{itemize}

\item [Feature adequacy]\nl\vspace{-1.6em} 
  \begin{itemize} 
  \item if $M \in \NT$ and $f = v$ is a feature of $M$ then, for each
    node $N$ in $\rI(M)$, either $v$ is an admissible value for the feature $f$
    in $N$ or $N$ does not contain the feature name $f$;
  \item if $M, N \in \NP$ both contain a feature $f$ with the same
    co-reference, then the values associated with $f$ in $\I(M)$ and
    $\I(N)$ are identical.
  \end{itemize}

\item [Node type adequacy]\nl\vspace{-1.6em} 
  \begin{itemize} 
  \item if $M \in \NP$ is an anchor with phonological form $phon$, then
    $PP(\I(M)) = [phon]$;
  \item if $M \in \NP$ is empty then $PP(\I(M)) = []$;
  \item if $M \in \NP$ is full then $PP(\I(M)) \not= []$.
  \end{itemize}

\item [Saturation]\nl\vspace{-1.6em} 
  \begin{itemize} 
  \item the multiset of polarities associated to a feature name $f$ in
    the set of nodes in $\rI(M)$ which contains the feature $f$ is
    globally saturated.
  \end{itemize}

\item [Minimality]\nl\vspace{-1.6em} 
  \begin{itemize} 
  \item ${\cal I}$ is surjective;
  \item if $M, N \in \NT$ and $M \gg N$ then there is $M' \in \rI(M)$
    and $N' \in \rI(N)$ such that $M' > N'$;
  \item if $M \in \NT$ and $f = v$ is a feature of $M$ then at least
    one node in $\rI(M)$ contains a feature with name $f$;
  \item if $M \in \NP$ is a leaf node with a non-empty phonological
    form $phon$, then $\rI(M)$ contains exactly one anchor node with
    phonological form $phon$.
  \end{itemize}
\end{description}

The four points defining minimality control the fact that ``nothing''
is added when the model is built. They respectively control the
absence of node creation, parenthood relation creation, feature
creation, and phonological form creation.

Note that there can be more than one interpretation function for a
given tree model.

\subsection{Polarized grammars}
\label{polarized_grammars}
An {\em interaction grammar} $\cal{G}$ is defined as a set of IPTDs.
The tree language defined by the grammar $\cal{G}$ is the set of
syntactic trees which are the models of a multiset of IPTDs from
$\cal{G}$. The string language defined by a grammar is the set of
phonological projections of the trees in the tree language.

We said that a syntactic tree $\T$ is a {\em parse tree} of a sentence
$\S$, that is a list of words $\S = w_1, \ldots w_n$ if:
\begin{itemize}
\item $\T$ is a model of some multiset of IPTDs from $\G$,
\item $PP(\T) = [w_1, \ldots, w_n]$.
\end{itemize}

An interaction grammar is said to be {\em lexicalized} if each IPTD
contains at least one anchor (an anchor is a leaf with
a non-empty phonological form).

An interaction grammar is said to be {\em strictly lexicalized} if each
IPTD contains exactly one anchor. In this case, the link with the
words of the language can be seen as a function which maps a word to
the subset of IPTDs which have this word as the phonological form of its
anchor. The grammar written so far for French is strictly lexicalized. 



\section{The expressivity of Interaction Grammars}
\label{sec:expressivity}
We present four aspects of IG that highlight their expressivity. We
illustrate these aspects with examples taken from our French IG
because it is the only IG which is fully implemented at the moment,
but there is no essential obstacle to use IG with other languages (an
English IG is being written).

\subsection{The use of polarities for pairing grammatical words}
\label{pairing}
In French, there are some grammatical words that are used in pairs:
\begin{itemize}
\item comparative,
  \french{plus $\ldots$ que} (more $\ldots$ than),
  \french{moins $\ldots$ que} (less $\ldots$ than),
  \french{si $\ldots$ que} (so $\ldots$ that), 
  \french{aussi $\ldots$ que} (as $\ldots$ as);

\item negation, 
  \french{ne $\ldots$ pas} (not), 
  \french{ne $\ldots$ rien} (nothing), 
  \french{ne $\ldots$ aucun} (no),
  \french{ne $\ldots$ personne} (nobody),
  $\ldots$;

\item coordinating words like 
  \french{soit $\ldots$ soit $\ldots$} (either $\ldots$ or),
  \french{ni $\ldots$ ni $\ldots$} (neither $\ldots$ nor),
  \french{ou $\ldots$ ou bien $\ldots$} (either $\ldots$ or).

\end{itemize}

The difficulty of modelling them is that their relative position in the sentence
is more or less free. For instance, here are examples that illustrate various
positions of the determiner \french{aucun} used with the particle \french{ne}:


\begin{examples}
\item
  \gll [Aucun] collègue [ne] parle à la femme de Jean.
  No colleague {} talks to the wife of John.
  \glt `No colleague talks to John's wife.'
  \glend
\item
  \gll Jean [ne] parle à la femme d' [aucun] collègue.
  John {} talks to the wife of no colleague.
  \glt `John talks to no colleague's wife.'
  \glend
\item
  \gll Le directeur dans [aucune] entreprise [ne] décide seul.
  The director in no compagny {} decides alone.
  \glt `The director in no compagny decides alone.'
  \glend
\item 
  \gll Jean [n'] est à la tête d' [aucune] entreprise.
  John {} is at the head of no compagny.
  \glt `John is at the head of no compagny.'
  \glend
\item \label{ne_aucun_e}
  \gll \agram Jean qui dirige [aucune] entreprise, [n]'est satisfait.
  John who heads no compagny, isn't satisfied. 
  \glt  \glend
\end{examples}

\begin{figure}
\begin{center}
\begin{tabular}{c|c}
\includegraphics[width=4cm]{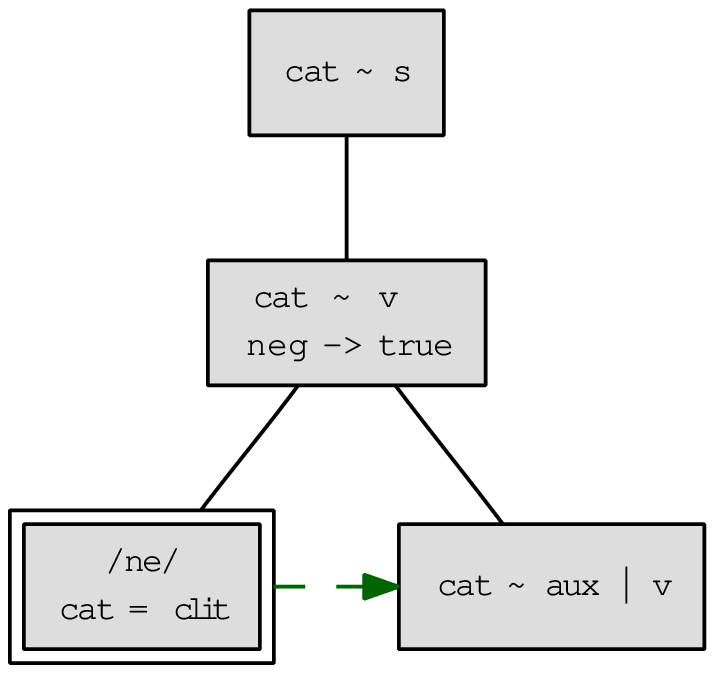}
&
\includegraphics[width=6cm]{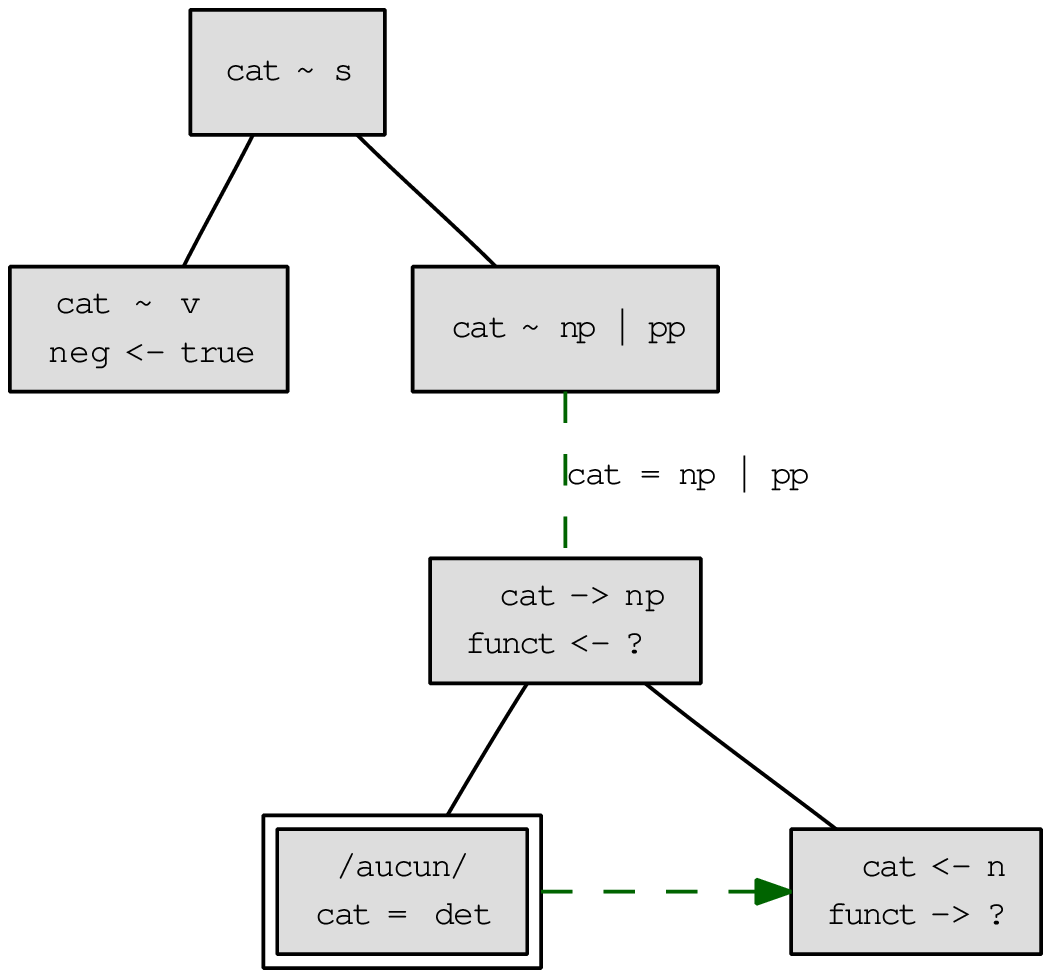}
\end{tabular}
\caption{IPTDs associated with the particle \french{ne} and the determiner
\french{aucun}}
\label{ne_aucun}
\end{center}
\end{figure}

The IPTDs from Figure~\ref{ne_aucun}, associated with the words \french{ne} and
\french{aucun}, allow all these sentences to be correctly parsed. The word
\french{ne} put a positive feature \fpos{neg}{true} on the maximal
projection of the verb that it modifies and this feature is neutralized by a
dual feature \fneg{neg}{true} provided by \french{aucun}. In its IPTD,
there is a constraint in the underspecified dominance relation that forbids the
acceptation of the sentence (\ref{ne_aucun_e}).

\subsection{Constrained dominance relations modelling long-distance
dependencies}
\label{long_distance}
Underspecified dominance relations are used to represent unbounded
dependencies and the feature structures that label these relations
allow for the expression of constraints on these dependencies, such as
barriers to extraction.

Relative pronouns, such as \french{qui} or \french{lequel}, give rise to unbounded
dependencies in series, a phenomenon that is called \emph{pied piping}.
Sentence~(\ref{ex_pied_piping_a}) is an example of pied piping.

\begin{examples}
\item \label{ex_pied_piping_a}
  \gll Jean [dans l' entreprise de {\bf qui}]  Marie sait que l' ingénieur travaille $\square$  est malade.
  John in the compagny of whom Mary knows that the engineer works {} is sick.
  \glt `John, in the compagny of whom Mary knows the engineer works, is sick.'
  \glend
\item \label{ex_pied_piping_b}
  \gll \agram Jean [dans l' entreprise de {\bf qui}] Marie qui travaille $\square$ le connaît est malade.
  John in the compagny of whom Mary who works {} knows it is sick.
  \glt  \glend
\item \label{ex_pied_piping_c}
  \gll \agram Jean [dans l' entreprise qui appartient à {\bf qui}] Marie travaille $\square$  est malade.
  John in the compagny which belongs to whom Mary works {} is sick.
  \glt  \glend
\end{examples}

\begin{figure}[h]
\begin{center}
\includegraphics[width=8cm]{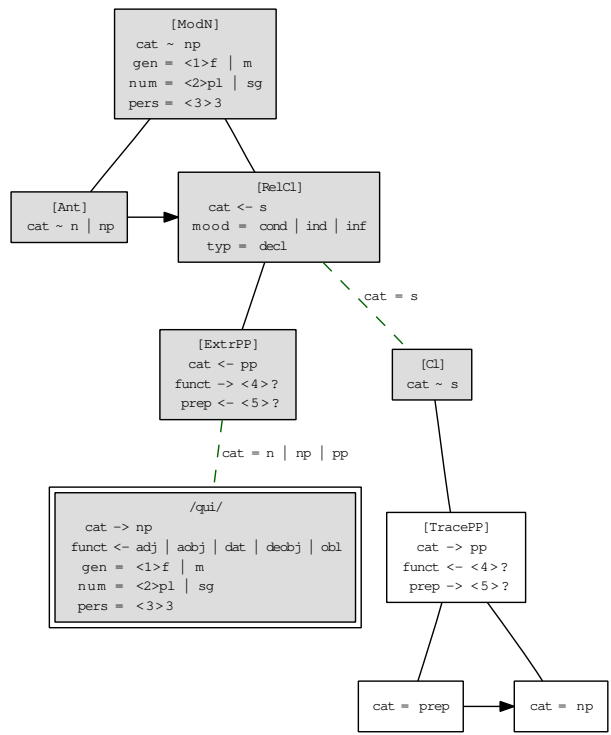}
\caption{IPTD associated with the relative pronoun \french{qui} used in an oblique
complement}
\label{qui}
\end{center}
\end{figure}

In example~(\ref{ex_pied_piping_a}), there is a first unbounded dependency
between the verb \french{travaille} and its extracted complement \french{dans
l'entreprise de qui}. The trace of the extracted complement is denoted by the
symbol $\square$.  This dependency is represented with an underspecified
dominance relation in the IPTD describing the syntactic behaviour of the relative
pronoun \french{qui} on figure~\ref{qui}.  The dominance relation links the
node [RelCl] representing the relative clause \french{[dans l'entreprise de {\bf
qui}]  Marie sait que l'ingénieur travaille $\square$} and the node [Cl]
representing the clause \french{que l'ingénieur travaille $\square$}, in which the
extracted prepositional phrase \french{dans l'entreprise de {\bf qui}} plays the
role of an oblique complement. The filtering feature structure labelling the
relation expresses that the path from [RelCl] to [Cl]  can only cross
a sequence of object clauses. This way, the sentence~(\ref{ex_pied_piping_b}) is
rejected because the dependency crosses a noun phrase, which violates the
constraint.

Inside the extracted prepositional phrase, there is a second unbounded
dependency between the head of the phrase and the relative pronoun \french{qui},
which can be embedded more or less deeply in the phrase. This dependency is also
represented on figure~\ref{qui} with an underspecified dominance relation. This
dominance relation links the [ExtrPP] node and the node representing the
relative pronoun \french{qui} and the associated filtering feature structure
expresses that the embedded constituents are only common nouns, noun phrases or
prepositional phrases. Finally, the sentence~(\ref{ex_pied_piping_c}) is
rejected.

\subsection{Adjunction of modifiers with virtual polarities}\label{adjunct}
\label{modifier}
In French, the position of adverbial complements in a sentence is relatively
free, as the following examples show:

\begin{examples}
\item
  \gll {\bf {\em Le soir}}, Jean {va rendre visite à} Marie.
  {At night}, John visits Mary. 
  \glt `At night, John visits Mary.'
  \glend
\item \gll Jean, {\bf {\em le soir}}, {va rendre visite à} Marie.
  John, {at night}, visits Mary. 
  \glt `At night, John visits Mary.'
  \glend
\item \gll Jean {va rendre visite} {\bf {\em le soir}} à Marie.
  John visits {at night} {} Mary. 
  \glt `John visits Mary at night.'
  \glend
\item \gll Jean {va rendre visite à} Marie {\bf {\em le soir}}.
  John visits Mary {at night}. 
  \glt `John visits Mary at night.'
  \glend
\end{examples}

These variants express different communicative intentions but the adverbial
complement \french{le soir} is a sentence modifier in all cases.
 
The virtual polarity \virtual\ was absent from the previous version of
IG~\cite{Per05}. Modifier adjunction was performed in the same way as in several
formalisms (CG, TAG) by adding a new level in the syntactic tree including the
modified constituent: instead of a node with a category X, we inserted a tree with
a root and two daughters;
 \begin{itemize}
 \item the root represents the constituent with the category X after modifier
adjunction;
 \item the first daughter represents the  constituent with the category X before
modifier adjunction;
 \item the second daughter represents the modifier itself.
 \end{itemize}
 Sometimes, this introduction of an additional level is justified, but
 most of the time it brings additional artificial complexity and
 ambiguity. Borrowing an idea from the system of black and white
 polarities of A.~Nasr~\cite{Nas95}, we have introduced the virtual
 polarity \virtual.  This polarity allows for the introduction of a
 modifier as an additional daughter of the node that it modifies
 without changing anything in the rest of the tree including the
 modified node.  Figure~\ref{qui} gives an example of an IPTD
 modelling a modifier: the relative pronoun \french{qui}, after
 combining with the relative clause that it introduces, provides a
 modifier of a noun phrase. The noun phrase to be modified is the
 antecedent of the relative pronoun, represented by node [Ant] and the
 noun phrase, after modification, is the root [ModN] of the IPTD.

\subsection{The challenge of coordination}
\label{coordination}
Even if we restrict ourselves to syntax, modelling coordination is a challenge.
First, there is no consensus about the analysis of the phenomenon in thslae
communauty of linguists~\cite{Carston03,Godard05}.
Then, whatever the chosen approach is, formalization encounters serious
obstacles. In particular, both Phrase Grammars and Dependency Grammars have
difficulties for modelling coordination of non-constituents.

J.~Le Roux and G.~Perrier propose to model coordination in IG with the
notion of polarity~\cite{Ler2006,Ler2007}. From this notion, they
define the interface of a PTD as the nodes that carry positive,
negative or virtual polarities. The interface characterizes the
ability of a phrase to interact with other phrases.  Two phrases can
be coordinated if the PTDs representing their syntactic structure
offer the same interface. Then, coordination consists in merging the
interfaces of the two PTDs. This merging needs to superpose several
positive or negative polarities and it also requires parse structure
to be DAGs rather than trees. Hence, the merge of two interfaces
cannot be modelled directly in IG and it is simulated in the PTD
associated with a coordination conjunction: this is divided into three
parts; two lower parts are used to saturate the interfaces of the
conjuncts and a higher part presents the common interface to the
outside.

With this principle, it is possible to parse the following sentences, which
illustrate different kinds of non-constituent coordination:

\begin{examples}
\item\label{coord_a}
  \gll Jean [boit du vin] et [mange du pain].
  John drinks {} wine and eats {} bread.
  \glt `John drinks wine and eats bread.'
  \glend
\item\label{coord_b} 
  \gll [Jean aime] mais [Marie déteste] la compétition.
  John likes but Mary dislikes {} competition.
  \glt `John likes but Mary dislikes competition.'
  \glend
\item\label{coord_c} 
\gll Jean donne [des fleurs à Marie] et [des bonbons à Pierre].
  John gives {} flowers to Mary and {} candies to Peter. 
  \glt `John gives flowers to Mary and candies to Peter.'
  \glend
\item\label{coord_d} 
  \gll La destruction [de la {gare routière} par les bombes] et [de la {gare ferroviaire} par les tanks] rend l' accès à la ville difficile.
  The destruction of the {bus station} by {} bombs and of the {railway station} by {} tanks makes {} access to the city difficult.
  \glt `The destruction of the bus station by bombs and of the railway station by tanks makes access to the city difficult.'
  \glend
\item\label{coord_e} 
  \gll Jean voit [sa soeur lundi] et [son frère mardi].
  John sees its sister {on monday} and its brother {on tuesday}.
  \glt `John sees its sister on monday and its brother on tuesday.'
  \glend
\item\label{coord_f} 
  \gll [Jean aime {le ski}] et [Marie  $\square$ la natation].
  John likes skiing and Mary {} swimming.
  \glt `John likes skiing and Mary likes swimming.'
  \glend

\end{examples}

Sentences (\ref{coord_a}) and (\ref{coord_a}) respectively illustrate left and
right node raising. Sentences (\ref{coord_c}) and (\ref{coord_d}) illustrate
coordination of argument clusters. Sentence (\ref{coord_e}) coordinates clusters
mixing arguments and adjuncts. Sentence (\ref{coord_f}) illustrates the
coordination of sentences with gaps. Here, the gap, which is represented by the
$\square$ symbol, corresponds to the elided verb \french{aime}.

\section{Comparison with other formalisms}
\label{sec:comparison}
Currently, there exists no linguistic formalisms that prevails over
the others. This means that the domain of natural language modelling
is still in an embryonic state and the congestion of the market is not
a good reason for not examining any new proposal. On the
contrary, the market is open.  But any new formalism has to show some
advantages with respect to the established ones in order to
survive. The challenge is to approximate linguistic generalities as
much as possible while remaining tractable. Remaining tractable means
being able to build large scale grammars and efficient parsers.  Under
this angle, the number of relevant formalisms is not that important:
among the most well known and largely used, there are LFG, HPSG, TAG
or CCG.  The comparison of IG with other formalisms will highlight
some of its strong features.

\subsection{Categorial Grammar}
The list of linguistic formalisms above mentions CCG (Combinatory Categorial
Grammars)~\cite{Ste00}. CCG are part of the CG family and since IG stems from
CG, it is natural to begin the comparative study with CG. 

IG shares with CG the fact that syntactic composition is based on the
resource sensitivity of natural languages, a property which is
built-in in both kinds of formalisms.  However, they differ in the
framework that they use. For this, we refer again to the distinction
between two approaches for syntax introduced by G.~Pullum and
B.~Scholtz~\cite{Pul01} and we can claim that CG uses a
generative-enumerative syntactic (GES) framework whereas IG uses a
model-theoretic syntactic (MTS) framework. In other words, CG derives
all acceptable sentences of a language from a finite set of axioms,
the lexicon, using a finite set of rewriting rules. IG associates
sentences with a set of constraints, which are solved to produce their
syntactic structures.

\cite{Per01} proposes a method for transposing grammars from the GES to the MTS
framework under some conditions. This method applies to CG and can be used to
compare IG with CG by putting them in the same MTS framework. The precise
description of such a translation goes beyond the goal of this article but we
give an outline of its output. 

To be more precise, let us focus on a particularly interesting member
of the CG family: CCG. The formalism of CCG is a very good compromise
between expressivity, simplicity and efficiency. At the same time, it
is able to model difficult linguistic phenomena, the most famous being
coordination~\cite{Ste85}, and it is used for parsing large corpora
with efficient polynomial algorithms and large scale
grammars~\cite{Hoc2003,Cla2004}.

If we use the method proposed in~\cite{Per01} to translate a particular CCG
in the MTS framework, we obtain a very specific IG with the following features
as output:
\begin{itemize}
\item Each syntactic type is translated into an IPTD with a particular
  shape.  Nodes are labelled with feature structures which contains
  only the \feat{cat} feature. The values of this feature are the
  atomic types of the CCG. Immediate dominance relations always go
  from nodes with a positive feature to nodes with a negative feature
  (possibly with intermediate nodes without labels). For large
  dominance relations, this is the contrary.
\item In the output IPTD, there are no precedence relations. Word order
  is controlled by a special feature \feat{phon}, which gives the
  phonological form of each node. This feature is neutral and takes
  its values from the monoid of the words of the language. We need to
  extend the system of IG feature values to allow the presence of
  variables inside terms representing \feat{phon} values.  These
  variables are used to model the sharing of unknown substrings of
  words by \feat{phon} values of different nodes.
\item Successful CCG derivations are translated into constructions of
  IPTD models. However, all valid IPTD models do not correspond to
  successful derivations, because the particular form of the
  combinatory rules imposes constraints to superposition. Conversely,
  in very rare cases, CCG derivations cannot be translated into
  constructions of IPTD models because of two rules: backward and
  forward crossed compositions. By allowing word permutation, these
  rules contradict the monotony of the MTS framework. A simple
  solution consists in discarding the two problematic rules and
  considering only a restriction of CCG.

\end{itemize}
Even if the translation of a CCG into an IG is not perfect, this
highlights the difference between the two formalisms. CCG can be
viewed as IG with additional constraints on the form of IPTDs and
superpositions. What is important, is that node merging is restricted
to pairs of nodes with dual \feat{cat} features. This has two
important consequences:

\begin{itemize}
\item It is not possible to express passive constraints on the environment of a
syntactic object, as we do in IG using nodes with virtual and neutral features.
\item The internal structure of an IPTD, that is its saturated nodes, is ignored
by CCG. The only thing that matters is its interface, that is its unsaturated nodes. 
\end{itemize}
The abstraction power that is expressed by this last remark is a source of
over-generation for CCG.
To limit over-generation,~\cite{Bal2003} have introduced modalities to control
the applicability of combinators rules. These modalities are specified in the
lexicon, so that the syntactic behaviour of a word can be more or less
constrained. The problem is that we cannot relativize these constraints with
respect to the environment in which the word can be situated. 
For instance, consider the following sentence:
\begin{example}\label{ccgexample}
Mary whom John met yesterday is my wife.
\end{example}
 In CCG, the relative pronoun \french{whom} provides
an object for the clause that it introduces on the right periphery of this
clause,  but the transitive verb \french{met} expects its object immediately on
its right. The way to solve this contradiction is to assign a modality to the
lexical entries of \french{met} and \french{yesterday}, which allow the permutation
of the object of \french{met} with \french{yesterday}.
But, doing this, we make the following sentence acceptable:
\begin{example}
* John met yesterday Mary.
\end{example}
IG does not present such an drawback, because \french{yesterday} is
taken as a sentence modifier and it is modelled according to the
method presented in subsection~\ref{adjunct}.

To summarize, multi-modal CCG limits over-generation but does not eliminate it.

\subsection{Dependency Grammars}
Like CG, Dependency Grammar (DG)~\cite{Niv2005} does not denote a unique
formalism but rather a family of formalisms. At the root of this family, there
is the concept of \emph{dependency}.
A dependency links two words in an asymmetrical manner: one word is the
\emph{r\'egissant} and the second word is the \emph{subordonn\'e}, according to
the terminology introduced by L. Tesni\`ere, the pioneer of DG~\cite{Tes59}. 

Even if there is no explicit notion of polarity in DG, this underlies
the notion of dependency. The potentiality of two words to establish a
dependency between themselves can be expressed by equipping the
\emph{r\'egissant} with a negative feature and the \emph{subordonn\'e}
with a positive feature, the two features having the same value, the
POS (part-of-speech) of the \emph{subordonn\'e} for instance.  This is
the general idea, which must be made more precise by examining the
different DG formalisms.  A key feature which differentiates DG
variants is the relationship between dependency structure and word
order.

Projective DG forbid cross-dependencies. They have interesting
computational properties and they can be easily translated into phrase
structure grammars, especially Adjukiewicz-Bar-Hillel (AB)
grammars~\cite{Bar60}. Since AB grammars can be viewed as CCG with
only two combinatory rules, forward and backward applications, the
consequence is that projective DG can be translated into IG following
the method presented above. This translation highlights the limits of
projective DG.  In fact, these are not expressive enough to represent
cross-dependencies or long-distance dependencies.

If we look at non projective DG, there is no formalism that has
reached sufficient maturity to be used for developing real
grammars. Nevertheless some works are promising and we propose to
focus on Generalized Categorial Dependency Grammar
(GCDG)~\cite{Dek2007}, which constitute a good compromise between
expressivity and complexity.

GCDG include two kinds of dependencies, thus giving birth to two independent formal
systems: 
\begin{itemize}
\item  projective dependencies are represented by AB grammars, slightly extended
to better take modifiers into account,
\item discontinuous dependencies are represented with polarities that
neutralize themselves in dual pairs.
\end{itemize}
A word that is able to govern another one in a discontinuous
dependency is equipped with a negative polarity typed by the category
of the \emph{subordonn\'e} and the \emph{subordonn\'e} is equipped
with the dual polarity.

This representation of discontinuous dependencies makes the comparison with IG
difficult. It is not possible to translate it in the framework of IG because it
has no simple relationship with dominance and precedence relations, which consider
phrases and not words. 
In IG, discontinuous dependencies are generally represented in the IPTD
associated with only one of the word responsible for the dependency, by means of
an underspecified dominance relation (see section~\ref{sec:expressivity}).

Another reason that makes the comparison between IG and GCDG difficult is that
there is no effective GCDG for any language. Nevertheless, we can make some
remarks. In GCDG, the iteration operator $\ast$ allows to represent modifiers by
sister adjunction as in IG. On the other hand, the hermetic separation between
the two kinds of dependencies does not allow to express that the same words
require a dependency when it does not matter if the dependency is projective or
discontinuous.

Because of the fine dependency structure that they propose, GCDG can contribute
to make clearer a controversial issue in DG, the analysis of grammatical
function words, but they will be confronted to syntactic constructions, which
remain problematic for all DG: coordination for instance.

\subsection{Unification Grammars}
The family of Unification Grammars (UG) includes all formalisms for which the
mechanism of unification between feature structures occupies a central position.
HPSG~\cite{Sag03} is the member of this family for which the idea is integrated
as completely as possible. The grammatical objects are typed feature
structures (grammatical rules, lexical entries and partial analysis structures)
and the only composition operation is unification.

From some angle, HPSG feature structures can be viewed as DAGs, in which edges
are labeled with feature names and leaves with atomic feature values. In this
way, unification appears as DAG superposition. As in IG, superposition gives
flexibility to HPSG and allows to represent sophisticated passive contexts of
syntactic constructions. 

The main difference is that the notion of unsaturated structure is not built-in
in the composition mechanism such as for IG with the notion of polarity.
However, this notion is present in some grammatical principles such as
the \emph{Valence Principle}.

Moreover, HPSG presents three important differences with respect to IG
\begin{itemize}
\item DAG are more expressive than trees. In this way, some phenomena are easier
to model with HPSG than with IG. For instance, factorization, which is specific
to coordination, is directly represented in HPSG~\cite{Mou2007}, whereas it must
be simulated in IG (see paragraph~\ref{coordination} and~\cite{Ler2006}).
\item Underspecification is more restricted in HPSG than in IG; it reduces to
the underspecification associated with unification. All dominance relations are
completely specified, so that unbounded dependencies are represented with
another mechanism: the \emph{slash feature}, the propagation of which allows to
mimic unbounded dependencies.
\item word order is not expressed by linear order between DAG nodes but with a
specific feature PHON.
\end{itemize}

Lexical Functional Grammar (LFG)~\cite{Bre2001} is another well known
member of the UG family, but because of their functional structures
paired with constituency structures, they are difficult to compare
with IG IPTDs. Nevertheless, presenting functional structures as
path equations allows the expression of a form of underspecification,
which is not present in HPSG but which exists in IG: the concept of
\emph{functional uncertainty} is similar to the IG notion of large
dominance, with the same possibility of constraining the dominance
path between nodes without determining its length.

Tree Adjoining Grammar (TAG)~\cite{Abe01} is often ranked in the UG
family, even if they are rather tree grammars but their use of
unification is more limited: contrarily to previous formalisms, it
cannot be used to superpose structures. Structures only combine by
adjunction, which greatly limits the expressivity of the formalism.
\section{Computational aspects}
\label{sec:leopar}

A question that arises naturally for a new formalism is its
complexity. The theoretical complexity is an important point but the
less formal notion of ``practical'' complexity is also crucial for
applications. The practical complexity can be thought as: ``how does
the formalism behave with real grammars and real sentences?''.

It is clear that IG is not as mature as the other formalisms presented in
the previous section. However, some theoretical and practical works
presented in this section give some insights about this question in
the IG framework.

The current work focuses on strictly lexicalized IG: the methods and
algorithms presented in this section apply to grammars where each IPTD
contains exactly one anchor. For such a grammar, we call {\em lexicon}
the function that maps each word to its corresponding set of IPTDs.
However, it is easy to transform any
lexicalized grammar into an equivalent strictly lexicalized grammar with
the mechanism used in section~\ref{pairing}.

In the particular case of strictly lexicalized grammar, the definition
of section~\ref{polarized_grammars} can be refomulated as follows. A
sentence $\S = w_1, \ldots w_n$ has a parse tree $\T$ iff there is an
ordered list of IPTDs $\P = [\P_1, \ldots, \P_n]$ such that:
\begin{itemize}
\item for all $1 \leq i \leq n$, $w_i$ is the phonological form of the anchor of
$\P_i$;
\item $\T$ is a model of the multiset $\{\P_1, \ldots, \P_n\}$;
\item $PP(\T) = [w_1, \ldots, w_n]$.
\end{itemize}

Hence, the parsing process can be divided in two steps:
first, select for each word of the sentence one of the IPTDs given by
the lexicon; then build a syntactic tree which is a model of the list
of IPTDs chosen in the first step. The choice of one IPTD for each word
of the sentence is called a {\em lexical selection}.

\subsection{Complexity}
The general parsing problem for IG is NP-complete, even if the grammar
is strictly lexicalized. It can be shown for instance with an encoding
of a fragment of linear logic (Intuitionistic Implicative Linear
Logic) in IG. Intuitively, the complexity has two sources:

\begin{itemize}
\item {\bf Lexical ambiguity}. In a lexicalized IG, each word of the
  lexicon can be associated to several IPTDs. Hence, the numbers of
  lexical selections for a given sentence grows exponentially with the
  number of words it contains.
\item {\bf Parsing ambiguity}. When a lexical selection is done, a
  model should be built for the corresponding list of IPTDs. Building
  a model is equivalent to finding a partition on the set of nodes of
  the IPTDs such that each node obtained by the mergings of nodes that
  are in the same subset of the partition are saturated. Once again,
  there is an exponential number of possible partitions.
\end{itemize}

The next two subsections address these two sources of ambiguity.  As
already mention above, we address the problem of practical
complexity. Hence, we are looking for algorithms which behave in an
interesting way for real NLP grammars. For instance, the formalism can
be used to define a grammar without any active polarity, but this is
clearly out of the IG ``spirit''. The methods described below are
designed for well-polarized grammars.

\subsection{Global filtering of lexical selections}
\label{filter}
In this section, we describe a method which is formalized in a
previous paper~\cite{Bon04} and we see how it applies to the IG
formalism. The idea is close to tagging, but it relies on more
precise syntactic descriptions than POS-tagging. Such methods are
sometimes called super-tagging~\cite{boullier2003}: we consider an
abstraction of our syntactic structures for which parsing is very
efficient even if this abstraction brings over-generation. The key
point is that a lexical selection which is not parsed in the abstract
level cannot be parsed in the former level and can be safely removed.

In IG, we consider as an abstract view of an IPTD the multiset of active
features present in the IPTD. Then, a lexical selection is valid in the
abstract level if the union of the multiset associated to IPTDs is
globally saturated.

The parsing at this abstract level is efficient because it can be done
using finite state automata (FSA). For each couple $(f,v)$ of a feature
name and a feature value, an acyclic automaton is build with IPTDs as
edges and integers as state: the integer in a state is the count of
polarities (positive counts for $1$ and negative for $-1$) for the
couple $(f,v)$ along every paths from initial state to the current
state. Finally, only lexical selections which end with a state
labelled with $0$ should be kept.

An automaton is built for each possible couple $(f,v)$, then a
FSA intersection of the set of automata describes the set of
lexical selections that are globally saturated.

The fact that feature values can be disjunction of atomic values in
IPTDs causes the automata to be non deterministic. We turn them into
deterministic ones using intervals of integers instead of integers in
states of the automaton.

When a grammar uses many polarized features, the method can be very
efficient and remove many bad lexical selections before the deep
parsing step. For instance, for the sentence~(\ref{ex:filter}) the number
of lexical selections reduces from 578\,340 to 115 (in 0.08s).

\begin{examples}\label{ex:filter}
\item 
  \gll L' ingénieur le présente à l' entreprise.
  The engineer him presents to the enterprise.
  \glt `The engineer presents him to the enterprise.'
  \glend
\end{examples}

The main drawback of the method is that the count of polarities is
global and does not depend on word order: any permutation of a
saturated lexical selection is still saturated. Some recent or ongoing
works try to apply some finer filters on automaton. In
\cite{leroux_bonfante}, a specialized filter is described dealing with
coordination for instance. For each IPTD for a symmetrical
coordination, this filter removes the IPTD if it is not possible to
find two sequences of IPTD on each side of the coordination with the
expected multiset of polarities.

\subsection{Deep parsing}
Deep parsing in IG is a constraint satisfaction problem. Given a
list of IPTDs, we have to find the set of models of the corresponding
multiset which respects the word order of the input sentence.

Three algorithms have been developed for deep parsing in IG:
\begin{description}
\item [Incremental] This algorithm scans the sentence word by word. An
  atomic step consists in chosing a couple of positive and
  negative features to superpose. In others words, an interpretation
  function is built step by step, guided by the {\bf saturation}
  property of models.
\item [CKY-like] The CKY-like algorithm, as the incremental one, tries
  to build the interpretation function step by step. The difference
  with the previous one is the way the sentence is scanned; it is done
  by filling a chart with partial parsing corresponding to sequence
  $[i,j]$ of consecutive words.
\item [Earley-like] This last algorithm tries to build at the same
  time the tree model and the interpretation function. It proceeds
  with a top-down/left-right building of the tree.
\end{description}

\subsubsection{Node merging}
The first two algorithms use the same atomic operation of node
merging. This operation takes as input a PTD $D$ and a couple of nodes
$(N_1,N_2)$; it returns a new PTD $D'$ which verifies that each model
of $D'$ is a model of $D$.

The model searching can be decomposed in small node merging steps
because of the following property: if the unsaturated PTD $D$ has a
model $T$ then there are two dual nodes $N_1$ and $N_2$ such that
$T$ is still a model of the PTD obtained by merging of $N_1$ and $N_2$ in $D$.

Technically, when two nodes are merged, some other constraint
propagation rules can be applied to the output description without
changing the set of models. For instance, if $M_1 > N_1$ and $M_2 >
N_2$ and $N_1$ is merged with $N_2$ then $M_1$ is necessarily merged
with $M_2$.

\subsubsection{The incremental algorithm}
As already said, there is an exponential number of possible choices of
couples of nodes to merge. The incremental algorithm tries to mimic
the human reading of a sentence and uses a notion of bound inspired by
psycholinguistics motivations to guide the parsing. This notion of
bound is used in a very similar spirit in Morrill's works~\cite{Mor00}.

The psycholinguistic hypothesis is that the reading uses only a small
memory to represent the already read part of the sentence. Hence, we
bound the number of unresolved dependencies that can be left open
while scanning the sentence. In our context, we bound the number of
active polarities. Then the algorithm uses a kind of shift/reduce
mechanism: we start with an empty PTD and then we used recursively the
two rules:

\begin{description}
\item [REDUCE] if the current PTD has a number of active polarities
  greater than the bound or if there is no more IPTD to add, then try the
  different ways to neutralize two dual active features;
\item [SHIFT] else, add the next IPTD to the current PTD.
\end{description}

In the \leopar\ implementation, the search space is controlled in the
REDUCE operation. Couples of active polarities are ordered in such a
way that multiple constructions of the same model which differ only by
a permutation on the neutralizations order are avoided.

\subsubsection{The CKY-like algorithm}
The well-known CKY parsing algorithm for CFG can be adapted to IG. The
basic idea is to focus on contiguous sequence of words and to use the
following informal rule:

A PTD for a sequence $[i,j]$ is obtained with a neutralization of two
dual features in two different PTDs for sequences $[i,k]$ and $[k+1,j]$.

This rule is used recursively to fill a chart. In the end, we consider
the PTDs obtained for the whole sentence and search for models: use
the REDUCE rule of the previous algorithm until there is no more
active polarity and second, build a totally ordered tree which is a
model of the saturated PTD obtained in the first step.

The advantages of this algorithm is that it does not depend on a bound
and that it is able to share more sub-parsing. The drawback is that it
is designed to find only models that follow some continuity conditions:
for instance, it is not able to find a model if neutralization arises
between $w_1$ and $w_3$ in a 3 words sentence. However, in our French
grammar, this condition is most of the time respected. But this
algorithm should be generalized in order to deal with other languages.

\subsubsection{The Earley-like algorithm}
Another algorithm inspired by the classical Earley parsing algorithm
for CFG has been developed for IG. The algorithm is described in
\cite{Ler2007,master_marchand}. It is being implemented in \leopar\
and the current version is not very efficient but we hope to improve
it for the next release.

There are two main difficulties to adapt this classical algorithm to
IG. First, when trying to build the tree model top-down, we have to
deal with large dominance relations. If the node $M$ is used to build
a node in the tree model and if $M >^* N$, then the node $N$ must be
used at any depth in the construction of the sub-tree rooted in $M$.
Our solution is to include in each item a set of nodes that must be
used in the subtree rooted at the current node. The other difficulty
is to deal with the fact that the daughters of a node are only
partially ordered in IPTDs and that we have to consider every total
ordering compatible with the partial order when building the tree
structure of the model.

\subsection{Implementation}
The IG formalism is implemented in a parser named \leopar. This
software contains several modules which are used in turn for sentence
parsing.
\begin{description}
\item[Tokenizer] a minimal tokenizer is included: it allows to deal
  with usual tokenization problems like contraction (for instance in
  French, the written word \french{au} should be understood as the
  contraction of the two words \french{à} and \french{le}). The
  tokenizer returns an acyclic graph to represent tokenization
  ambiguities.
\item[Lexer] a flexible system of linguistic resources description is
  used in \leopar. Several levels of description can be used to
  described various linguistic information: morphological,
  syntactical,\ldots. Unanchored IPTDs are read in an XML format
  produced by XMG~\cite{Duc04} (an external tool which provides a high
  level language to build large coverage grammars). The anchoring
  mechanism is controlled by the notion of interface: 
  \begin{itemize}
  \item each description tree of the unanchored grammar is associated
    with a feature structure called interface;
  \item each word is linked to a set of usages: a usage is a feature
    structure which describes the morphological and syntactical
    properties of a word;
  \item if an interface $I(T)$ of a tree description $T$ unifies with
    a word usage $U$ associated with a word $w$: then an IPTD $T'$ is
    produced from $T$ with $w$ as phonological form.
  \end{itemize}
  The lexer outputs an acyclic graph which edges are labelled by IPTDs.

\item[Filter] this stage implements the global filtering of lexical
  selections presented above (subsection~\ref{filter}). It takes as
  input the acyclic graph given by the lexer and returns another
  acyclic graph which paths are the lexical selections kept by the
  filtering process.

\item[Deep parser] the final stage is the building of a set of models
  for the acyclic graph given by the previous stage. Implemented
  algorithms are adapted to deal with the sharing given by the graph
  representation of the ambiguity in the output of the filtering
  process.
\end{description}

The whole system can be used either with commands or through an
interface. In the interface, an interactive mode is available. The
user can choose a path in the automaton given by the filter stage and
then choose couple of nodes to merge: this interactive mode is very
useful in grammar testing/debugging.

\section{Conclusion}
In this paper, we focused on a formal presentation of IG, highlighting their
originality with their ability to express various and sophisticated linguistic
phenomena. We left both Language-Theoretic properties and implementation
aspects of IG aside, as they need to be studied for themselves.

One of our fundamental ideas is to combine theory and practice.  The
formalism of IG is implemented in the \leopar\ parser in the same form
as it is described in this paper. In this way, it can be validated
experimentally. To use \leopar\ on large corpora, we need
resources. There exists a French IG with a relatively large coverage
\cite{Per2007}, which is usable with a lexicon independent of the IG
formalism~\cite{Gar2005}. There exists a lexicon with a large coverage
available in the format required by the grammar: the
\lefff~\cite{Sagot2006}. The \lefff\ contains about 500\,000 inflected
forms corresponding, among others, to 6\,800 verb lemmas, 37\,600 nominal
lemmas and 10\,000 adjectival lemmas. With the \lefff\ and the French IG,
\leopar\ is on the way of parsing real corpora.

The formalism is not definitively fixed and the forward and backward motion
between theory and practice is important to improve it step by step.
Among the questions to be studied in a deeper way, there are:

\begin{description}
\item[\em the form of the syntactic structure of a sentence:]
  phenomena such as coordination or dislocation show that the notion
  of syntactic tree is too limited to express the complexity of the
  syntactic structure of sentences; structures as directed acyclic
  graphs fit in better with these phenomena;
\item[\em the enrichment of the feature dependencies:] dependencies
  between features are frequent in linguistic constructions but they
  cannot be represented in a compact way in the current version of IG;
  all cases have to be enumerated, which is very costly; it seems not
  to be a difficult problem to enrich the feature system in order to
  integrate these dependencies.
\end{description}

The paper is restricted to the syntactic level of natural languages
but syntax cannot be modelled without any idea of the semantic level
and of the interface between the two levels;~\cite{Per05} presents a
first proposal for the extension of IG to the semantic level but we
can envisage other approaches using existing semantic formalisms such
as MRS~\cite{Cop2005} or CLLS~\cite{Egg2001}.

\bibliography{ig}
